\documentclass[amsmath,amssymb,notitlepage]{revtex4-1}

\usepackage{graphicx}

\usepackage{bm}

\newcommand{\ket}[1]{| #1 \rangle}
\newcommand{\Ket}[1]{\bigl| #1 \bigr\rangle}
\newcommand{\bra}[1]{\langle #1 |}
\newcommand{\cla}{\mathcal}
\newcommand{\Op}[1]{\hat{#1}}
\newcommand{\Eq}[1]{Eq.~(\ref{#1})}
\newcommand{\Id}{\openone}

\begin{document}
\renewcommand{\_}{{\tt\textunderscore}} 

\title{Cloud-Assisted Contracted Simulation of Quantum Chains }

\author{Alexander Yu.\ Vlasov} 
\email[Electronic mails: ]{a\_y\_vlasov@yahoo.com, qubeat@mail.ru}
\affiliation{
 P.V.~Ramzaev Research Institute of Radiation Hygiene\\ 
 8 Mira Street, Saint Petersburg 197101, Russia
}

\date{16 October 2019} 

\begin{abstract}
 The work discusses validation of properties of quantum circuits with many qubits using
 non-universal set of quantum gates ensuring possibility of effective simulation on classical computer.
 An understanding analogy between different models of quantum chains is suggested for clarification.
 An example with IBM Q Experience cloud platform and Qiskit framework is discussed finally.
\end{abstract}

\maketitle

\section{\label{sec:intro} Introduction}

A question about compliance with the model of {\em scalable gate-based quantum computations} 
desirable for generally known algorithms from BQP ({\em bounded-error quantum 
polynomial time\/}) complexity class \cite{BQP} encounters certain difficulties already for not very big 
amount of qubits, because of problems with direct verification of results even using 
modern supercomputers.

The discussions about ``quantum supremacy'' \cite{supr} milestone rather emphasize such 
a controversy, because it requires comparison between some quantum devices and state-of-art 
classical simulators \cite{simsupr}. Different methods to address such a problem 
could be suggested. Well-known example is random sampling using appropriate (universal) 
set of quantum gates together with statistical analysis of final states of qubits \cite{blue}. 
An alternative approach is looking for ``quantum agreement'' with a specific quantum circuits 
with restricted (non-universal) sets of gates generating entanglement of hundreds or even thousands 
qubits, but effectively modelled by classical computers.

One possible example is specific version of {\em logarithmic space bounded
quantum computations} implemented by so-called {\em matchgates} \cite{val,log}. 
In such a case some non-universal quantum circuit with $d$ qubits can be 
{\em ``contracted''} into universal one with only $l = \lceil \log_2(d) \rceil$ qubits.
Simplified version of such approach using illustrative example with quantum chains is discussed
below. 

\section{\label{sec:qchain} Quantum Chains}

\begin{figure}[h]
\center{\includegraphics[width=.55\linewidth]{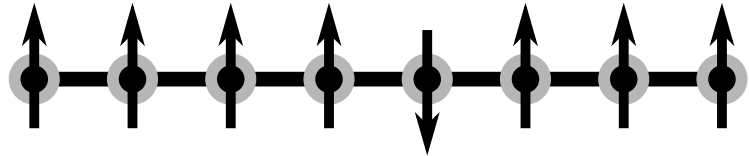}}
\caption{Qubit (spin) chain.}
\label{Fig:quchain}
\end{figure}

A correspondence between two models is used in presented work.
The first one is a chain with $d$ qubits (Figure~\ref{Fig:quchain}). 
The space of states for such a model has dimension $2^d$.

\begin{figure}[h]
\center{\includegraphics[width=.55\linewidth]{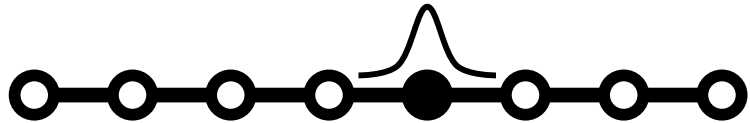}}
\caption{Quantum scalar chain.}
\label{Fig:schain}
\end{figure}

The second model is a quantum scalar chain with $d$ nodes (Figure~\ref{Fig:schain})
with only $d$ states that can be considered as a single ``qudit,'' 
that is, in turn, could be `contracted' (`compressed') into $l = \lceil\log_2(d)\rceil$ qubits. 

Let us consider quantum system with Hamiltonian
\begin{equation}\label{H}
\Op{\cla{H}} =  
 \sum_{j=0}^{d-2} \frac{\lambda_j}{2} (\Op\sigma^x_j \Op\sigma^x_{j+1} + \Op\sigma^y_j \Op\sigma^y_{j+1}) 
+ \sum_{j=0}^{d-2} \frac{\chi_j}{2} (\Op\sigma^y_j \Op\sigma^{x\vphantom{y}}_{j+1} 
                                     - \Op\sigma^{x\vphantom{y}}_j \Op\sigma^y_{j+1}) 
+ \sum_{k=0}^{d-1} \mu_k \Op\sigma^z_k.
\end{equation}
It should be mentioned, that the Hamiltonian \Eq{H} does not include forth order terms required 
for generation of universal set of quantum gates ({\em cf} analogue expression in Ref.~\cite{blue}). 

It may be checked directly, that \Eq{H} commutes with `number operator'
\begin{equation}\label{opN}
 \Op{\cla N} = \sum_{k=0}^{d-1}\frac{\Op{\Id} - \Op\sigma^z_k}{2}
 \end{equation} 
Thus, number ${\cla N}$ of units in computational basis is {\em conserved}
by quantum gates $\Op{\cla G} = \exp(-i \Op{\cla H} \Delta t)$ 
generated by Hamiltonians \Eq{H}. Here $\Delta t$ is 
time interval of application for given $\Op{\cla{H}}$ 
and the {\em natural system of units} with $\hbar=1$ is used in all equations.

The states with $\cla N = 1$ simply correspond 
to $d$ basic states of quantum scalar chain with respect to map
\begin{equation} \label{embone}
 \ket{k} \mapsto \Ket{\underbrace{0\ldots0}_k 1 \underbrace{0\ldots0}_{d-k-1}}, \quad
 k = 0,\ldots,d-1. 
\end{equation}
For qubit chain conservation of 
$\cla N$ conforms to {\em restricted case} of 
matchgate circuits \cite{TD2,AV18} and it is represented below by two-qubit gates 
such as $\Op{\cla{M}}$ in \Eq{mgu} or \Eq{mu3}. 

The term `matchgate' was introduced in \cite{val} for a quantum two-gate
of special form \Eq{mgvu} recollected below.
The considered model is also naturally 
represented using relation between {\em $Spin(2d)$ groups} and 
orthogonal transformations in dimension $2d$ \cite{AV18}.
In general, conservation of $\cla N$ is not mandatory, but it is discussed 
elsewhere \cite{TD2,AV18}.

Similar approach with {\it compressed \sl quantum computation} was tested on 
{\em IBM Q Experience\/} cloud platform \cite{compr}.
In such a case 5-qubits quantum chip was used for simulation of quantum Ising
chain with $2^5=32$ spins and for testing was used correspondence between 
$2$ and $2^2=4$ qubits with rather pessimistic results for current error level.

The term {\it contracted \sl quantum simulation} is chosen here because, from the one hand, 
exponentially smaller (`contracted') model is used for testing. On the other hand,
it provides informal reference to idea of reliable {\em design-by-contract} \cite{contr}
with natural tests (`contracts') for appropriate functionality. 
It may be useful for testing both quantum chips and 
classical simulators of quantum computer. 

The modelling was performed by author with {\em IBM Q Experience\/} {\tt Qiskit} \cite{qiskit} framework 
providing common environment for work with a few real quantum chips and simulation 
both on hight performance computer (HPC) in the cloud and personal computer (PC). 
Discussed model with chains is included in the community tutorials for {\tt Qiskit} 
\cite{chain-tut} and it is revisited in the next section.
More recent updates and extensions may be found in separate repository 
for quantum chain models \cite{quchain}. 

\section{\label{sec:sim} Quantum Walk Simulation}
\subsection{\label{subsec:walk} Discrete-time quantum walks}

\begin{figure}[htb]
\center{\includegraphics[width=.55\linewidth]{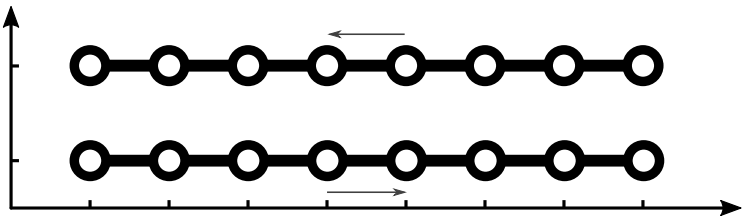}}
\caption{Space of states for coined quantum walk}
\label{Fig:quwalk}
\end{figure}

Model with continuous evolution described by Hamiltonian (\ref{H}) is not adapted
for implementation with quantum circuits. However, model of {\em coined quantum walks}
after appropriate reformulation for qubit chain also may be described using
similar approach \cite{AV18} and it can be used here with the similar purposes.

Let us start with usual model of {\em discrete-time quantum walks} for 
scalar quantum chains with further reformulation to quantum circuit model  
using correspondence \Eq{embone}.
{\em Coined quantum walk} \cite{Kem03} is naturally defined for composite quantum system 
with a chain $\ket{k}$, $k=0,\ldots,n-1$ and coin $\ket{c}$, $c = 0,1$.
The basic states of such a system can be expressed as $\ket{c}\ket{k}$ and
the dimension of space of states is $d=2n$. 
The coin and chain states are depicted on Figure~\ref{Fig:quwalk} along 
vertical and horizontal axes respectively.

\begin{figure}[htb]
\center{\includegraphics[width=.55\linewidth]{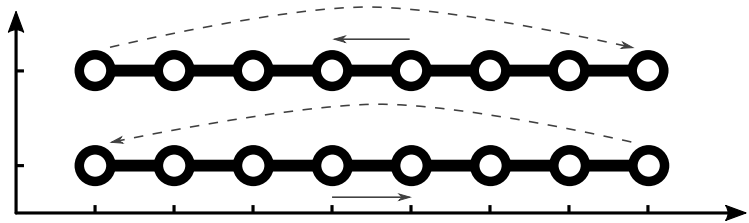}}
\caption{Periodic boundary conditions}
\label{Fig:quwalkcycl}
\end{figure}

Let us consider operators of right and left shift on the chain $\Op{\cla R}$ and $\Op{\cla L}$
together with an operator acting on composite system
\begin{equation}\label{OpS}
 \Op{\cla B} = \ket{0}\bra{0} \otimes \Op{\cla R} + \ket{1}\bra{1} \otimes \Op{\cla L}.
\end{equation}
The operator $\Op{\cla B}$ (`quantum bot' \cite{qubot}) could be considered as an example of 
{\em conditional quantum dynamics} \cite{cond95} with chain as a target and coin as control. 
For simplest case without superposition
of coin states it applies operators $\Op{\cla R}$ or $\Op{\cla L}$ to chain
for states of coin $\ket{0}$ or $\ket{1}$ respectively. For finite chains 
the periodic boundary conditions may be considered first for simplicity, 
see Figure~\ref{Fig:quwalkcycl}.

Coined quantum walk has more complex dynamics due to additional `coin toss' operator
$\Op{\cla C}$ acting on the control space. 
The standard choice for $\Op{\cla C}$ is Hadamard coin 
\begin{equation}\label{HCoin}
\Op{\cla C}_{\rm H} = \frac{\sqrt{2}}{2}%
\left(\begin{array}{rr}
  1 & 1 \\ 1 & -1
\end{array}\right)
\end{equation}
or balanced coin
\begin{equation}\label{bCoin}
\Op{\cla C}_{\rm b} = \frac{\sqrt{2}}{2}%
\begin{pmatrix}
  1 & i \\ i & 1
\end{pmatrix}.
\end{equation}
Taking into account such operator the single step of quantum walk can be expressed as
composition of $\Op{\cla B}$ and coin toss operator $\Op{\cla C}$
\begin{equation}
 \Op{\cla W} = (\Op{\cla C} \otimes \Op{\Id}) \, \Op{\cla B} ,
\end{equation}
where $\Op{\Id}$ is the identity operator on a chain and
$\Op{\cla C}$ is a coin toss operator such as \Eq{HCoin} or \Eq{bCoin}.

\begin{figure}[htb]
\center{\includegraphics[width=.55\linewidth]{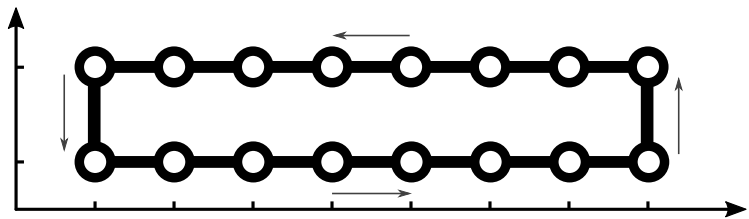}}
\caption{Reflecting boundary conditions}
\label{Fig:quwalkrefl}
\end{figure}

The cyclic (periodic) boundary conditions may be not very convenient for implementation with
neighbouring nodes mapped into qubit chain with nearest-neighbour quantum gates discussed below.
The operator $\Op{\cla B}$ can be modified for reflecting boundary conditions corresponding
to change of direction due to `flip' on the ends of chains, 
see Figure~\ref{Fig:quwalkrefl}.

\begin{figure}[htb]
\center{\includegraphics[width=.5\linewidth]{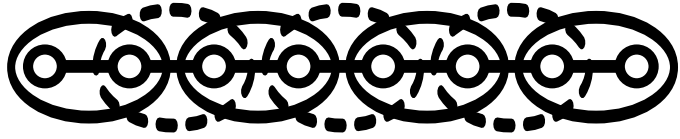}}
\caption{Partition for staggered quantum walk}
\label{Fig:qupart}
\end{figure}

For restriction of each operator to neighbouring nodes such a model can be represented
as so-called {\em staggered quantum walk} \cite{stag} on a chain with $d=2n$ nodes using correspondence
\begin{equation}\label{map2x1}
\ket{c}\ket{k} \longleftrightarrow \ket{2k+c}; \qquad k=0,\ldots,d-1; \quad c = 0,1.
\end{equation}
Let us consider such a chain with partitions depicted on a Figure~\ref{Fig:qupart} 
and a transformation produced by alternating swaps with pairs of nodes 
from first and second partition outlined by solid and dashed ellipses 
respectively.
 
Let us express swap of two nodes using Pauli matrix
\begin{equation}
\Op X = \Op\sigma_x = \begin{pmatrix}
  0 & 1 \\ 1 & 0
\end{pmatrix}.
\end{equation}
In such a case swap of all pairs in the first partition corresponds to operator 
\begin{equation}\label{partS1}
\Op{\cla B}_1 = \Op X_{(0,1)}\Op X_{(2,3)}\cdots \Op X_{(d-2,d-1)},
\end{equation}
there each operator in the expression swaps two nodes with indexes
shown in brackets. 
An analogue expression for the second partition is
\begin{equation}\label{partS2}
\Op{\cla B}_2 = \Op X_{(1,2)}\Op X_{(3,4)}\cdots \Op X_{(d-3,d-2)}.
\end{equation}
It may be checked directly, that with respect to map 
\Eq{map2x1} the composition of $\Op{\cla B}_1$ and $\Op{\cla B}_2$
implements operator $\Op{\cla B}$ for reflecting boundary condition.

The coin toss operator for a model of staggered quantum walk can be implemented
by application of the $\Op{\cla C}$ to each pair of nodes from the first partition
and may be expressed as
\begin{equation}\label{partC1}
 \Op{\cla C}_1 = \Op{\cla C}_{(0,1)}\Op{\cla C}_{(2,3)}\cdots \Op{\cla C}_{(d-2,d-1)}.
\end{equation}
In such a way the staggered walk is represented by composition of three operators
$\Op{\cla C}_1$, $\Op{\cla B}_1$ and $\Op{\cla B}_2$ acting only on neighbouring nodes.
The operators $\Op{\cla C}$ and $\Op{\cla B}_1$ act on the same pairs of nodes
and expression may be simplified by modification of coin toss operator
\begin{equation}\label{xcoin}
 \Op{\cla C}' = \Op{\cla C}\Op{X}
\end{equation}
with straightforward action on the first partition
\begin{equation}\label{partC'1}
 \Op{\cla C}'_1 = \Op{\cla C}'_{(0,1)}\Op{\cla C}'_{(2,3)}\cdots \Op{\cla C}'_{(d-2,d-1)}.
 \tag{\ref{partC1}$'$}
\end{equation}
For example with Hadamard coin modified operator is
\begin{equation}\label{xHCoin}
\Op{\cla C}'_{\rm H} = \Op{\cla C}_H\Op{X} = \frac{\sqrt{2}}{2}%
\left(\begin{array}{rr}
  1 & 1 \\ -1 & 1
\end{array}\right).
\end{equation} 
Thus, staggered walk is represented by composition of operators \Eq{partC'1} and \Eq{partS2}
\begin{equation}\label{Wstag}
 \Op{\cla W}_s =\Op{\cla C}'_1\, \Op{\cla B}^{}_2.
\end{equation}

\subsection{\label{subsec:chain} Modelling of qubit chain} 

The staggered walk on the chain can be simply implemented by {\tt Python} program
without any special libraries for simulation of quantum circuits, but analogue
model with qubit chain is using {\tt Qiskit}. Let us consider map \Eq{embone} 
introduced earlier to work with qubit chain. Operators acting on neighbouring 
nodes in such a case correspond to special case of matchgates.

Let us consider two unitary operators $\Op{u}$ and $\Op{v}$ 
represented by $2\times 2$ matrices with equal determinants 
\begin{equation}\label{opvu}
 \Op{v} = \begin{pmatrix}
  v_{00} & v_{01} \\ v_{10} & v_{11}
\end{pmatrix},\quad
 \Op{u} = \begin{pmatrix}
  u_{00} & u_{01} \\ u_{10} & u_{11}
\end{pmatrix},\quad \det \Op{v}=\det{\Op{u}}
\end{equation}
By definition the matchgate is two-gate on near neighbour qubits expressed as  $4 \times 4$ matrix
produced from elements \Eq{opvu} of $\Op{v}$ and $\Op{u}$ 
\begin{equation}\label{mgvu}
 \Op{\cla{M}}_{v,u} = \begin{pmatrix}
  v_{00} & 0 & 0 & v_{01} \\
  0 & u_{00} & u_{01} & 0  \\
  0 & u_{10} & u_{11} & 0 \\
  v_{10} & 0 & 0 & v_{11}
\end{pmatrix}.
\end{equation}

Let us consider special case with $\Op{v} = \Op\Id$ and $\det\Op{u} = 1$
\begin{equation}\label{mgu}
 \Op{\cla{M}}_u = \begin{pmatrix}
  1 & 0 & 0 & 0 \\
  0 & u_{00} & u_{01} & 0  \\
  0 & u_{10} & u_{11} & 0 \\
  0 & 0 & 0 & 1
\end{pmatrix}.
\end{equation}
Such matchgates are two-gates with non-trivial action only for
superpositions with states $\ket{01}$ and $\ket{10}$ of neighbouring qubits. 
It may be generated by Hamiltonians \Eq{H} with terms corresponding to considered 
pair of qubits. With respect to map \Eq{embone} it is equivalent with operator $\Op{u}$ 
acting on two neighbouring nodes.

For Hadamard coin `swapped' operator $\Op{\cla C}'_{\rm H}$ \Eq{xHCoin} has unit 
determinant and can be directly used for construction of matchgate
\begin{equation}\label{mgXC}
 \cla{\Op M_C'} = \begin{pmatrix}
  1 & 0 & 0 & 0 \\
  0 & 1/\sqrt{2} & 1/\sqrt{2} & 0  \\
  0 & -1/\sqrt{2} & 1/\sqrt{2} & 0 \\
  0 & 0 & 0 & 1
\end{pmatrix}.
\end{equation}
Such a gates should be applied to all pair of qubits in the first partition similarly
with \Eq{partC'1}

There is some subtlety, because swap operator {\em does not} have unit determinant
$\det(\Op{X}) = -1$ and operator $i\Op{X}$ with unit determinant was
used instead. Thus, two qubit-gates for a single swap is modified as
\begin{equation}\label{mgiX}
 \Op{\cla{M}}_{iX} = \begin{pmatrix}
  1 & 0 & 0 & 0 \\
  0 & 0 & i & 0  \\
  0 & i & 0 & 0 \\
  0 & 0 & 0 & 1
\end{pmatrix}.
\end{equation}
Application of such a gate for all pairs of qubits in the second partition is
analogue of \Eq{partS2}.

Single-qubit gates in {\tt Qiskit} and {\em quantum assembly language} 
{\tt OpenQASM} \cite{OpenQASM} are parametrized using three angles 
\begin{equation}\label{u3}
\Op{\cla U}(\theta,\phi,\lambda) = \begin{pmatrix}
e^{-i(\phi+\lambda)/2}\cos(\theta/2) & -e^{-i(\phi-\lambda)/2}\sin(\theta/2) \\
e^{i(\phi-\lambda)/2}\sin(\theta/2) & e^{i(\phi+\lambda)/2}\cos(\theta/2)
\end{pmatrix}.
\end{equation} 
With similar parametrization $\Op{\cla{M}}_u$ \Eq{mgu} can be rewritten
\begin{equation}\label{mu3}
\Op{\cla M}(\theta,\phi,\lambda) = \begin{pmatrix}
1 & 0 & 0 & 0 \\
0 & e^{-i(\phi+\lambda)/2}\cos(\theta/2) & -e^{-i(\phi-\lambda)/2}\sin(\theta/2) & 0 \\
0 & e^{i(\phi-\lambda)/2}\sin(\theta/2) & e^{i(\phi+\lambda)/2}\cos(\theta/2) & 0\\
0 & 0 & 0 & 1
\end{pmatrix}.
\end{equation} 
Now gates \Eq{mgXC} and \Eq{mgiX} can be rewritten
\begin{equation}\label{mupar}
\cla{\Op M_C'} = \Op{\cla M}(-\pi/2,0,0), \quad
 \Op{\cla{M}}_{iX} = \Op{\cla M}(\pi,\pi,0)
\end{equation}

With {\tt OpenQASM} notation the two-qubit gate \Eq{mu3} 
can be expressed as a sequence of single-qubit gates \Eq{u3} 
(denoted as {\tt U}) and {\em controlled NOT} gates 
(denoted as {\tt cx}).
{\samepage
\begin{quote}
\begin{verbatim}
M(theta, phi, lambda) a,b {
  cx a,b;
  U(0,0,(lambda-phi)/2) a;
  cx b,a;
  U(-theta/2,0,-(phi+lambda)/2) a;
  cx b,a;
  U(theta/2,phi,0) a;
  cx a,b;
}
\end{verbatim}
\end{quote}
}

Here variables in brackets are parameters and {\tt a}, {\tt b} are indexes of qubits.
Thus, such a function can be applied using parameters from \Eq{mupar} 
to necessary pairs of qubits. {\tt Qiskit} uses analogue approach with definition
of function in {\tt Python} language.

\end{document}